# 2GeV SUPERCONDUCTING MUON LINAC

Milorad Popovic

*Fermi National Accelerator Laboratory* [1]
*Batavia, IL 60510, USA*

*Abstract*

A muon collider as well as a neutrino factory requires a large number of muons with a kinetic energy of 50GeV or more. Muon survival demands a high gradient linac. The large transverse and longitudinal emittance of the muon beam coming from a muon cooling system implies the need for a large acceptance, acceleration system. These two requirements point clearly to a linac based on superconducting technology. The design of a 2GeV Superconducting muon Linac based on computer programs developed at LANL will be presented. The design is based on the technology available today or components that will be avaible in the very near future.

## 1 INTRODUCTION

Considerable interest has developed in the possibility of a high-energy, high luminosity μ–μ collider[1] and neutrino factory, and a multi-laboratory collaboration has been formed to study these concepts. The large number of muons needed for a collider, or for a muon storage ring, requires a high-intensity proton source for π-production, a high-acceptance π–μ decay channel, a μ-cooling system, a rapid acceleration system and a high-luminosity collider ring or muon storage ring with long straight decay sections. Presently, the baseline design for the acceleration system assumes a high-gradient straight linac followed by two recirculating linacs. The final energy is 50GeV and a six- month study at FNAL[2] has determined that the straight linac should have a final energy around 2GeV. The final energy of the first linac is dictated by the assumption that the next linac should be a recirculating one like the electron linac at Jefferson Lab. Recirculation saves money, and as the FNAL six-month study has shown, muon cooling and muon acceleration are the major cost drivers for a Neutrino Factory. In this paper the design of a high-gradient, straight linac for the muons will be presented. It is assumed that the muon beam that comes from the muon cooling system at 15 Hz rate has an energy of 200MeV and is bunched at 200MHz. Beam pulse contains $4 \times 10^{12}$ muons which are distributed in four pulses trains, 160ns long and 300ns apart. The transverse rms beam size is 55mm with an unnormalized rms beam emittance of $5 \times 10^4$ mm-mrad. The beam has an energy spread of +/- 40 MeV and the 200MHz bunches have a total length of 70cm.

| Beam | Value |
|---|---|
| Repetition Rate | 15 Hz |
| Beam RF Character. | 200MHz |
| Number of Muons | $4 \times 10^{12}$ |
| Kinetic Energy | 200MeV |
| Full Energy Spread | +/- 40MeV |
| Un. Emittance RMS | $5 \times 10^4$ mm-mrad |
| Transverse Beam Size | 55 mm (RMS) |
| Bunch Length | 70cm(total) |

Table 1

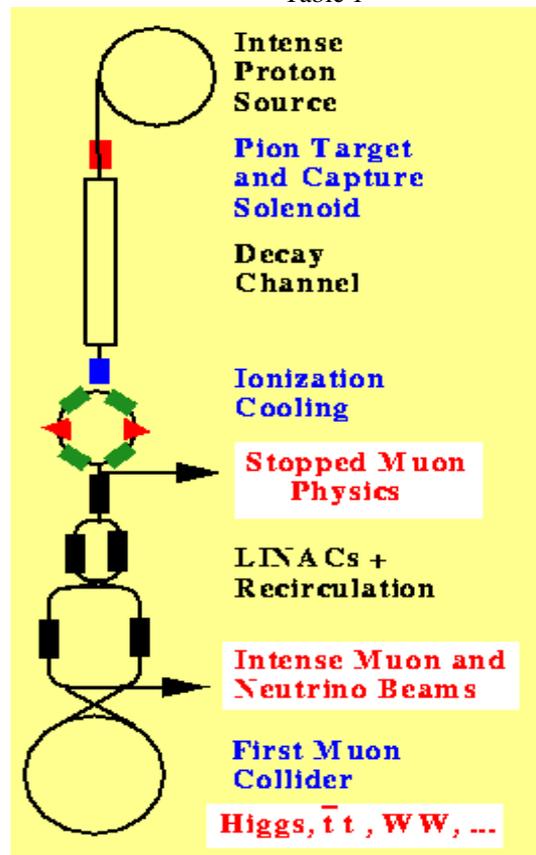

Figure 1.

[1]This work is supported by the U.S. Dept. of Energy through the Univ. Research Association under contract DE-AC35-89ER40486

These beam parameters force us to have a linac with an accelerating frequency of 200MHz or a lower multiple of it. Table 1 summarizes the parameters of the beam entering the linac. The large beam peak current (~1Amp) and short muon lifetime lead us to high gradient acceleration and high peak beam power. Figure 1 shows the survival rate of muons during acceleration from 100MeV to 2GeV, with an average gradient of 5MeV/m along the linac.

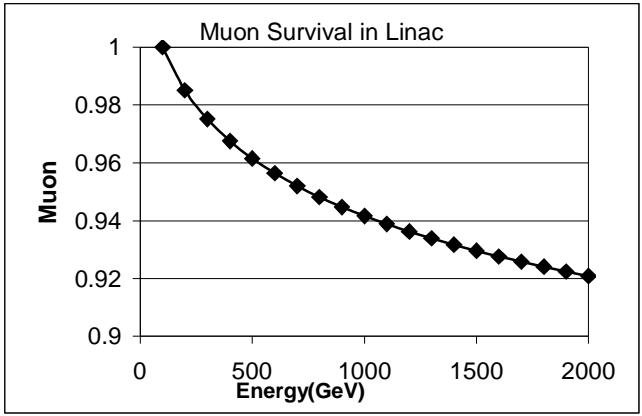

Figure 2.

The large transverse beam size points to a linac based on bell-shaped superconducting cavities.

## 2 LINAC DESIGN

The muon beam coming from the cooling system, fixes the frequency of the linac to be a multiple of the 200MHz. The beam emittance forces us to use cavities with a bore radius larger than 25 cm. The large peak current in short pulses and the long beam-off time suggest a long filling- time and acceleration based on the cavity's stored energy. The cavity's stored energy U is

$$U = gE^2\lambda^2$$

where, $\lambda$ is the RF wavelength, E is the electric field on axis and g is a proportionality factor that depends on the geometry of the cavity. The long filling time,

$$\tau = \frac{2Q_L}{\omega}$$

$\tau<30$ms, requires the loaded $Q_L$ to be less than $2\times10^7$.
Recent technology development suggests that acceleration gradient of 15MV/m at 200MHz is achievable and that a power coupler for 500kW will be available in the near future. All these puts constrains on the choice of design geometrical beta of 0.95 was chosen, with three cells per cavity. The geometrical beta of 0.95 is somewhat soft. An arbitrary choice of any value in the range of .95 to 0.98 will work with a change of less than 1% in the transit time factor. The cavity bore radius was chosen to be 27 cm, and the cavity beam pipe is one meter from each end of the cell. The cavity shape was calculated using so-called ELLFISH program [3]. The calculated stored energy is 930 Joule at 15MV/meter. The whole linac is designed with a constant energy gain of 21MeV per cavity, and each cryomodule contains one three-cell cavity. There is 17.5 cm of warm drift-space between the cryomodule and warm quad. The quad length is 35cm with quad gradient of 410Gauss/cm. A simple FODO lattice was chosen with a half-length of 4.26 meters. Figure 2 shows half of a 3-cell superconducting cavity for 201 MHz with bore radius of 27cm.

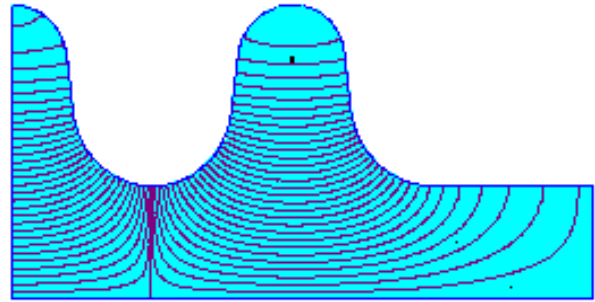

Figure 3.

The choice of a constant acceleration of 21MeV per cavity is somewhat arbitrary but it is consistent with today's achievable peak electric fields, limits on power couplers, and the requirement that the energy taken by the beam is only a few percent of the stored energy in the cavity. This will produce an energy droop between the head and tail of the beam pulse of only a few percent, well bellow the energy spread that is characteristic of the incoming beam .The beam phases for each cell as a function of beam energy is plotted in figure 2. The design choice of cavity phase represents a compromise between operation on the crest of the RF waveform to provide the maximal acceleration and operation earlier than the crest to provides longitudinal focusing. We start with a phase of 40 degrees at the entrance to the linac and ramp the phase linearly with the energy up to six degrees. The choice of 40 degrees at the entrance is a result of matching to the cooling section. The phase at the entrance and exit cell is calculated by the code PARMILA[4] based on the input and exit energy of the design particle in the cavity. The inputs to the design code are the desired energy gain per cavity and the phase of the design particle at the entrance to the linac. One may ask if does operation at a geometric beta far from the design velocity causes a large longitudinal emittance growth due to a large cell to cell phase variation through the cells of the first few cavities. The beam dynamics simulation showed no visible effect on the longitudinal emittance. This is perhaps not surprising for a beam with so large an energy spread as our muon beam.

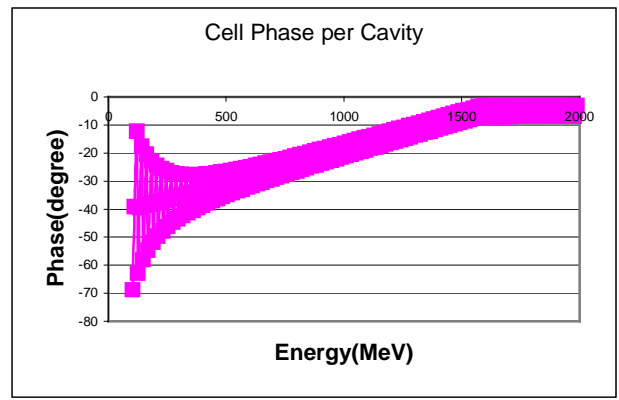

Figure 4.

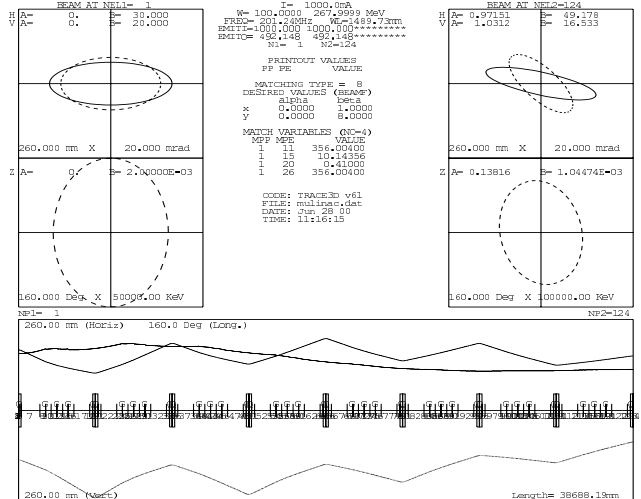

Figure 5.

Figure 5., shows Trace3D[5] run. The beam is matched into the linac, and first eight RF sections are shown.

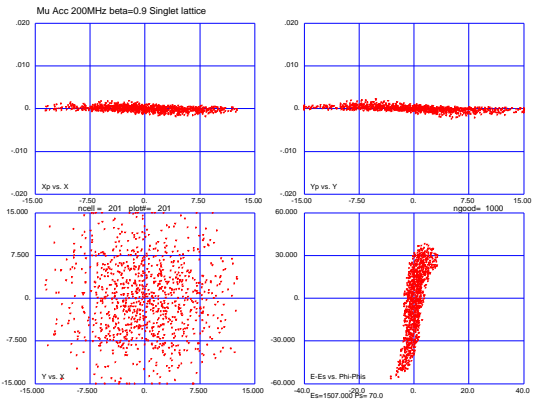

Figure 6.
Figure 6. shows x-xp, y-yp, x-y and angle-energy phase space plots for 500 particles pushed through the linac.

To accelerate beam to 2GeV, we need 90 cavities. The total length of the linac is 436 meters. We use conventional tubes as the RF source. One possibility is to use RCA Type 7835 tube, which is used for most of RF proton linacs in the US. This tube can deliver up to 5MW and can be used to power up to eight cavities in this linac.

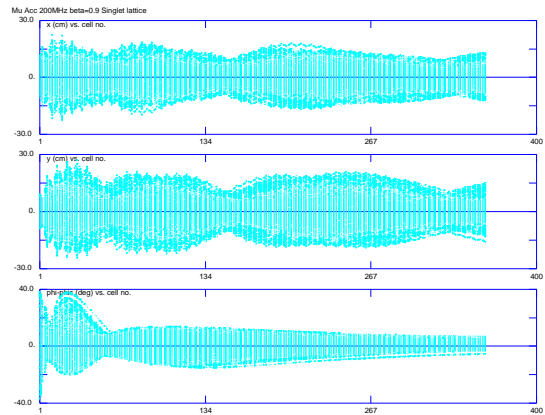

Figure 7

Figure 7. shows the beam profiles in x, y, and phase planes along the linac. The input beam parameters are in the usual Trace3D units, five times the RMS unnormalized beam emittance, and $\sqrt{5}$ times the beam RMS sizes

## CONCLUSIONS

Assuming today's technology[6] and hoping for modest advances in the field of superconducting RF a design of a 2GeV Superconducting linac is presented. The study shows that fast and efficient muon acceleration is possible and, based on the cost estimates of planned super- conducting proton linac[7], cost of this linac does not look out of reach in near future. One advantage of supercondacting linacs is the fact that energy upgrade is somehow build in characteristic and can be used either to shorten the linac or to increase injection energy in first recirculating linac. .